\documentstyle[multicol,aps,prl,epsf]{revtex}

\begin{document}

\draft

\title{
Superconductivity and spin correlation in organic conductors: \\
a quantum Monte Carlo study
}

\author{
Kazuhiko Kuroki and Hideo Aoki
}
\address{Department of Physics, University of Tokyo, Hongo,
Tokyo 113-0033, Japan}

\date{\today}

\maketitle

\begin{abstract}
The $d$-wave pairing correlations along with spin correlation are
calculated with quantum Monte Carlo method 
for the two-dimensional Hubbard model on 
lattice structures representing organic superconductors 
$\kappa$-(BEDT-TTF)$_2$X and (TMTSF)$_2$X.
In both cases the pairing correlations for
superconducting order parameters with nodes are found to be enhanced.
The symmetry and the enhancement of the pairing 
is systematically correlated with the spin structure factor, 
suggesting a spin-fluctuation mediated pairing.  
We have further found that, as we deform the Fermi surface to make
the system approach the half-filled square lattice, 
the coherence of the pairing saturates while 
the local pairing amplitude continues to increase.
\end{abstract}

\medskip


\begin{multicols}{2}
\narrowtext

\newpage
It is a great theoretical challenge to explore whether
the superconductivity in `exotic materials' 
such as the cuprates, heavy fermion systems,
and (at least some) organic conductors can 
be encompassed into a single class, i.e., superconductivity 
mediated by spin-fluctuations.
For more than a decade, the possibility of spin-fluctuation-mediated 
pairings has in fact been investigated intensively both theoretically and 
experimentally for heavy fermion systems\cite{hfrev} and 
high $T_C$ cuprates.\cite{htcrev} 
There, one important sign of spin-fluctuation-mediated pairing 
has been the superconducting gap with nodes.

There is now a body of accumulating experimental evidence that 
organic superconductors\cite{orgrev} $\kappa$-(BEDT-TTF)$_2$X
\cite{Mayaffre,Kanoda1,Kanoda2,Soto,Nakazawa}
and (TMTSF)$_2$X\cite{Takigawa}
also have non-$s$-wave gap like in heavy fermion or cuprate systems.
The unconventional pairing, along with the proximity of superconductivity to
antiferromagnetic or spin density wave state, 
suggest that pairing may be mediated by spin-fluctuation.
The pseudogap-like behavior of $1/(T_1T)$
above $T_C$ 
in $\kappa$-(BEDT-TTF)$_2$X \cite{Mayaffre,Kanoda2,Soto,Kawamoto} 
is also reminiscent of a similar behavior in the underdoped high $T_C$ 
cuprates, which would suggest that 
electron correlation may play an important role 
there.\cite{McKenzie}

Theoretically, a simplest many-body Hamiltonian to incorporate 
the electron correlation is the Hubbard model. 
Some analytical calculations have supported spin-fluctuation 
mediated pairing in the Hubbard model on lattices representing
$\kappa$-(BEDT-TTF)$_2$X\cite{Schmalian,Kino,VD,KM,Louati} 
or (TMTSF)$_2$X\cite{Shimahara}.
However, numerical evidences supporting such a possibility 
have yet to come.

Thus the purpose of the present paper is to explore numerically
the pairing correlation 
in the Hubbard model for lattices representing
$\kappa$-(BEDT-TTF)$_2$X and (TMTSF)$_2$X, 
with a special attention payed to whether 
the pairing is linked with the behavior of the spin correlation.  
Comparing the results for the two cases, we find that 
the symmetry of the pairing is indeed determined by the 
the peak position of the spin structure factor, 
\begin{figure}
\begin{center}
\leavevmode\epsfysize=40mm \epsfbox{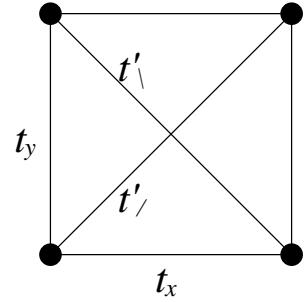}
\caption{The hopping parameters considered.
}
\label{Hamiltonian}
\end{center}
\end{figure}
which supports spin-fluctuation mediated pairing. 
We further deform the Fermi surface from the case corresponding to  
$\kappa$-(BEDT-TTF)$_2$X to find that the coherence of pairs 
saturates as the system approaches 
the half-filled square lattice (i.e., where, at least for large enough $U$, 
the system becomes a Mott insulator with antiferromagnetic order).
Here we adopt the ground-state, canonical-ensemble 
quantum Monte Carlo (QMC) method.   

We start from the Hubbard model on a two-dimensional lattice
(Fig.\ref{Hamiltonian}),
\begin{eqnarray}
{\cal H}&=&-\sum_{x,y,\sigma}
\left[t_x(c_{x,y,\sigma}^\dagger c_{x+1,y,\sigma})
+t_y(c_{x,y,\sigma}^\dagger c_{x,y+1,\sigma})\right.\nonumber\\
&+&\left.t'_{/}(c_{x,y,\sigma}^\dagger c_{x+1,y+1,\sigma})
+t'_{\backslash}(c_{x,y,\sigma}^\dagger c_{x-1,y+1,\sigma})+{\rm h.c.}
\right]\nonumber\\
&+&U\sum_{x,y} n_{x,y,\uparrow}n_{x,y,\downarrow}.
\end{eqnarray}
Here, $(x,y)$ is the coordinate of a site with 
the lattice constant taken to be unity, and 
periodic boundary condition is assumed.

While QMC method has been widely used to investigate the Hubbard model 
on the 2D square lattice, 
enhanced pairing correlation had eluded detection.  
Recently, however, the present authors identified this as coming 
form the fact that the pair-scattering processes that produce 
superconductivity has a very small energy scale of $O(0.01t)$ or less
\cite{KA1,KA2}: 
if one takes a closed shell condition (a parameter set with 
no ground state degeneracy for $U=0$), 
the energy gap between the highest occupied levels (HOLs) and the
lowest unoccupied levels (LULs) for $U=0$ is as large as $\sim O(0.1t)$ for 
tractable system sizes, so that the effect of the low energy pair scatterings 
would be smeared out. On the other hand, QMC calculations 
with an open-shell condition, in which the effect of low-energy pair 
scatterings is expected to be incorporated, 
suffer from numerical difficulties such as the negative-sign problem.
An open-shell configuration is ill-conditioned also in the sense that 
there are a finite number of levels 
within an infinitesimal distance from the Fermi energy, so that 
one can suspect an enhancement of pairing correlations, if any, 
may be due to such an effect.  

We have circumvented the problem 
by taking slightly different values of $t_x$ and
$t_y$ to lift the degeneracy between wave numbers $(k_1,k_2)$ and $(k_2,k_1)$,
and put the Fermi level in between those levels. This way 
we can prevent the low-energy pair scattering processes from being masked 
and at the same time take a closed shell configuration.
If we take typically $t_y/t_x=0.999$, which gives
HOL-LUL gap $\Delta$ of $<0.01t$, and $U/t=1$, 
we can achieve convergence with respect to the projection 
imaginary time $\tau$ in the QMC 
without running into serious sign problem.\cite{tech}

We calculate the pairing correlation functions,
\begin{eqnarray}
P(r)=\sum_{|\Delta x|+|\Delta y|=r} 
&&\langle O^\dagger (x+\Delta x,y+\Delta y) O (x,y)\nonumber\\
&&+O(x+\Delta x,y+\Delta y) O^\dagger (x,y)\rangle
\label{paircorr}
\end{eqnarray}
with 
\begin{equation}
O(x,y)=\sum_{\delta_x,\delta_y,\sigma}
\sigma(c_{x,y,\sigma} c_{x+\delta_x,y,-\sigma}
-c_{x,y,\sigma} c_{x,y+\delta_y,-\sigma}), 
\end{equation}
which includes the conventional $d_{x^2-y^2}$ symmetry 
(with $\delta_x=\delta_y=1$, corresponding to 
the order parameter proportional to $f({\bf k})\equiv\cos(k_x)-\cos(k_y)$
in $k$-space), but written here in a general form.  
Hereafter we define $r\equiv |\Delta x|+|\Delta y|$
as the real space distance.
To give an above-mentioned insight, we also calculate the spin 
structure factor,
\begin{equation}
S({\bf q})=
\frac{1}{N}\sum_{i,j}e^{i{\bf q}\cdot ({\bf r}_i-{\bf r}_j)}
\langle {\bf S}_i\cdot {\bf S}_j\rangle. 
\end{equation}

\begin{figure}
\begin{center}
\leavevmode\epsfysize=60mm \epsfbox{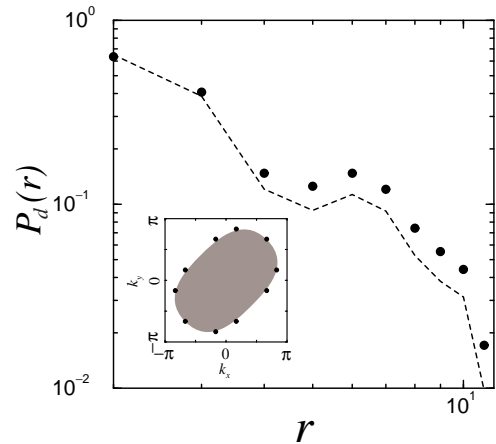}
\caption{
QMC result for the d$_{x^2-y^2}$ pairing correlation as a function of 
real space distance 
for 134 electrons in $12\times 12$ sites $(n=0.93)$ with $t_x=1$,
$t_y=0.999$, $t'_{\backslash}=0.70$, $t'_/=-0.11$,
$U=1$ (solid circles) and $U=0$ (dashed line).
The inset shows the HOLs and LULs within
0.01 to the Fermi energy. 
}
\label{bedtd-wave}
\end{center}
\end{figure}

\begin{figure}
\begin{center}
\leavevmode\epsfysize=50mm \epsfbox{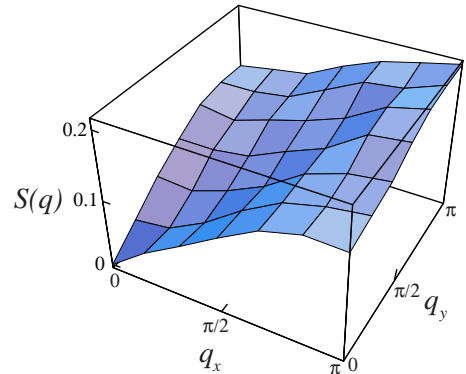}
\caption{
QMC result for the spin structure factor as a function of the 
wave vector. The parameters are the same as in Fig.\protect\ref{bedtd-wave}.
}
\label{bedtsdw}
\end{center}
\end{figure}

We first look into the case where the Fermi surface, represented by
HOLs and LULs, resembles those of $\kappa$-(BEDT-TTF)$_2$X materials.
It has been proposed that $\kappa$-(BEDT-TTF)$_2$X can be modeled by
a single band Hamiltonian on an anisotropic triangular lattice with
$t_x=t_y$ {\LARGE$_{\stackrel{>}{\widetilde{\:\:\:\:}}}$} 
$t'_\backslash\gg t'_/$ and $n=1$ (half-filled).
\cite{Tamura,Caulfield,KF}
This is because (i) the dimerization of the BEDT-TTF molecules is so 
strong that we can consider a dimer as a single site
as far as the low-energy excitations are concerned, and (ii) 
the hopping parameter in the $c$ direction alternates 
between two slightly different values, 
but the difference is small enough to be neglected as a 
first approximation.

First we take 134 electrons in a $12\times 12$ (band filling 
$n=0.93)$ lattice with 
$t_x=1$, $t_y=0.999$, $t'_\backslash=0.70$, $t'_/=-0.11$.
We have taken $t_y=0.999$ for the slight lift of degeneracy mentioned above.
The reason for taking small but negative $t'_/$ 
and the slight deviation from $n=1$ (to which the $\kappa$-(BEDT-TTF)$_2$X
system corresponds) is 
to distribute HOLs and LULs on the Fermi surface as
uniformly as possible.
In Fig.\ref{bedtd-wave}, 
we show the result for the $d_{x^2-y^2}$ 
pairing correlation $P_{x^2-y^2}(r)$ as a function of the 
real space distance $r$ for $U=1$.  
It can be seen that the pairing correlation 
for $U=1$ is enhanced over that for $U=0$,
especially at large distances.

If we look at the spin structure factor in $k$-space in Fig.\ref{bedtsdw}, 
a broad peak around $(\pi,\pi)$ is seen.
Such an antiferromagnetic spin-fluctuation enhances 
repulsive (in momentum-space) pair scattering processes  with 
momentum transfer $(\pi,\pi)$.
Repulsive pair scattering processes from $\sim\pm(0,\pi)$ to 
$\sim\pm(\pi,0)$ (and vice versa) 
favor a superconducting order parameter 
that takes the maximum of its absolute value with different signs 
around these two areas.
Thus, the enhancement of the $d_{x^2-y^2}$ pairing correlation 
along with the peak of $S({\bf q})$ at $(\pi,\pi)$ are 
consistent with the spin-fluctuation mediated pairing.

\begin{figure}
\begin{center}
\leavevmode\epsfysize=60mm \epsfbox{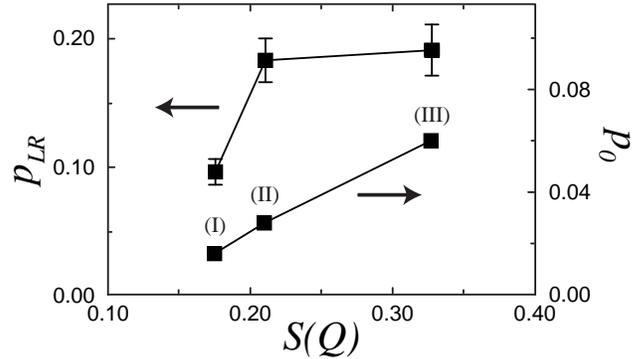}
\caption{
A plot for $p_{LR}$ and $p_0$ against 
$S({\bf Q})$ with ${\bf Q}=(\pi,\pi)$ for parameter sets
(I) $n=0.96$ $t_y=1.001$, $t'_{\backslash}=1.00$, $t'_{/}=0$,
(II) $n=0.93$, $t_y=0.999$, $t'_{\backslash}=0.70$, $t'_{/}=-0.11$ 
(corresponding to Fig.\protect\ref{bedtd-wave}), 
(III) $n=0.94$, $t_y=0.999$, $t'_{\backslash}=0.20$, $t'_{/}=0$,
all with $12\times 12$ sites, $t_x=1$, and $U=1$.}
\label{sdw-sup}
\end{center}
\end{figure}

We have further investigated the link between 
$d_{x^2-y^2}$ pairing correlation and the spin correlation
systematically by deforming the Fermi surface. 
In Fig.\ref{sdw-sup}, the enhancement of the pairing correlation
is plotted against $S({\bf Q})$ with ${\bf Q}=(\pi,\pi)$ 
for three cases I, II, and III. 
The main difference among the three cases 
is in the value of $t'_\backslash$ as given in the caption, 
while other slight differences are 
due to technical reasons. Namely,
we are going from the isotropic
triangular lattice (I) to 
a case close the square lattice (III) 
via case (II) (corresponding to Fig.\ref{bedtd-wave}) 
for $n$ fixed around unity. 
$S({\bf Q})$ increases as we approach the half-filled square lattice.
In Fig.\ref{sdw-sup}, the pairing is probed from two quantities: 
the pairing correlation (measured from 
the noninteracting case) summed over large distances, 
$p_{LR}\equiv\sum_{r\geq 4}\{[P_d(r)]_{U=1}-[P_d(r)]_{U=0}\}$, 
and the local amplitude of pairing, 
$p_0\equiv[P_d(0)]_{U=1}-[P_d(0)]_{U=0}$.

Fig.\ref{sdw-sup} shows that $p_0$ 
grows hand in hand with $S({\bf Q})$, endorsing the 
spin-fluctuation-mediated formation of $d_{x^2-y^2}$ pairs.
On the other hand, $p_{LR}$, 
the enhancement of the long-range part of $P_d(r)$, 
which measures the coherence of the pairs, also grows with $S({\bf Q})$ 
as we go from (I) to (II), but it
{\rm saturates} between (II) and (III),
suggesting that the coherence does not necessarily grow with
the local pair amplitude.

\begin{figure}
\begin{center}
\leavevmode\epsfysize=75mm \epsfbox{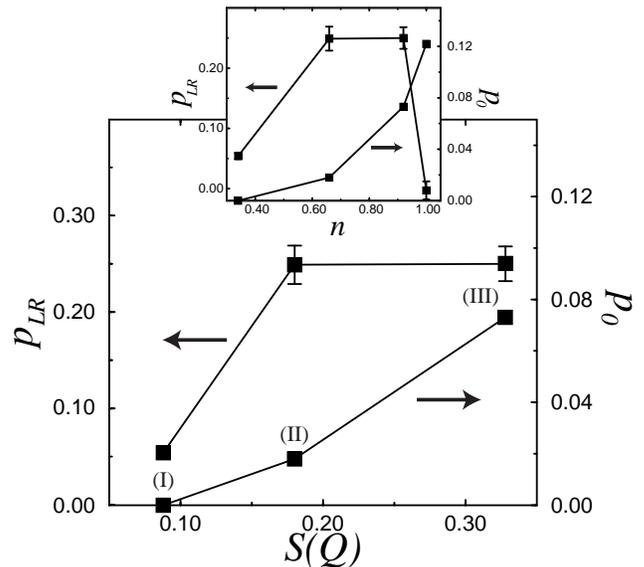}
\caption{
A plot similar to Fig.\protect\ref{sdw-sup} for a $10\times 10 $
square lattice
with (I) 34 $(n=0.34)$ (II) 66 $(n=0.66)$ (III) 92 $(n=0.92)$
electrons.
The hopping parameters are fixed at $t_x=1$, $t_y=0.999$, 
$t'_/=t'_\backslash=0$.
The inset shows the band filling dependence of $p_0$ and $p_{LR}$,
including the results for half-filling ($n=1$).
}
\label{sq-sdw-sup}
\end{center}
\end{figure}

To show that such a correlation between the $d$-wave pairing 
and the spin correlations is not accidental for the cases studied above,
we move on to 
the square lattice ($t_/=t_\backslash=0$) for various values of 
the band filling $n$.  
In Fig.\ref{sq-sdw-sup}, we show a plot similar to Fig.\ref{sdw-sup} for
the square lattice, along with the plot against $n$ in the inset.  
For intermediate densities, $p_{LR}$,
$p_0$, and $S({\bf Q})$ all grow as $n$ approaches the half-filling ($n=1$).
If we come too close to half-filling, however, 
$p_{LR}$ becomes saturated, while $S({\bf Q})$ and $p_0$ 
keeps growing. This is even more clearly seen 
at exactly half-filling, where $p_{LR}$ is strongly 
suppressed (inset of Fig.\ref{sq-sdw-sup}), while
$p_0$ continues to grow.
The result that, when $S({\bf Q})$ becomes too large, 
the pairs are formed but their coherence ($p_{LR}$) 
stops to grow unlike their local amplitude $p_0$ 
might have some relevance to the normal-state pseudogap behavior observed 
close to the superconducting-antiferromagnetic boundary
in $\kappa$-(BEDT-TTF)$_2$X as well as in the underdoped high $T_C$ cuprates.

\begin{figure}
\begin{center}
\leavevmode\epsfysize=110mm \epsfbox{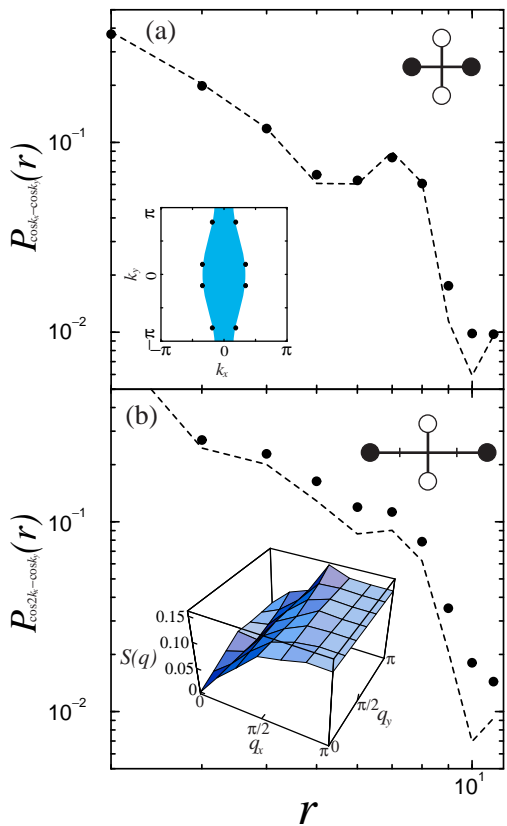}
\caption{
A plot similar to Fig.\protect\ref{bedtd-wave}
for 72 electrons in $12\times 12$ sites $(n=0.5)$ with $t_x=1$,
$t_y=0.212$, $t'_{\backslash}=t'_{/}=0$. 
Pairing correlations for $f({\bf k})=\cos k_x-\cos k_y$ (a)
(ordinary $d_{x^2-y^2}$), 
and for $f({\bf k})=\cos 2k_x-\cos k_y$ (b).
The upper inset in each figure schematically depicts 
the pairing ($\bullet$: $-$, $\circ$: $+$) in real space. 
The lower inset in (a) shows HOLs and LULs within 0.01 to the Fermi energy, 
while the spin structure factor is displayed in (b).
}
\label{tmtsfd-wave}
\end{center}
\end{figure}

In order to confirm spin-fluctuation mediated pairing 
to a wider extent, we next explore the case where the spin structure factor
is not peaked around $(\pi,\pi)$, by studying 
the case with a Fermi surface representing (TMTSF)$_2$X. 
If we neglect the weak dimerization along the stacking
direction, these materials may be modeled by a
single band Hamiltonian on a strongly anisotropic two-dimensional lattice
with $t_x/t_y\sim O(0.1)$ $(t'_\backslash=t'_/=0)$ and $n=0.5$ 
(quarter-filled).\cite{Shimahara,Yamaji} 
Here we take 72 particles in $12\times 12$ $(n=0.5)$, 
$t_x=1$, and $t_y=0.212$, for which the Fermi surface is as depicted in the 
lower inset of Fig.\ref{tmtsfd-wave}(a). 

For this Fermi surface consisting of 
warped parallel lines around $k_x\pm \pi/4$,
the nesting vector is $(\pi/2,\pi)$, so that 
the spin structure factor (lower inset of Fig.\ref{tmtsfd-wave}(b)) 
has a peak there.\cite{incomme} 
Then, pair scattering processes from $\sim\pm(0,\pi)$ to 
$\sim\pm(\pi/2,0)$ (and vice versa) would be enhanced, which 
favors a superconducting order parameter 
proportional to $f({\bf k})=\cos(2k_x)-\cos(k_y)$ 
($\delta_x=2$, $\delta_y=1$ in Eq.(\ref{paircorr}))
rather than the ordinary 
$d_{x^2-y^2}$ pairing with $f({\bf k})=\cos(k_x)-\cos(k_y)$. 
In fact, the possibility of such a pairing was pointed out in 
an RPA calculation.\cite{Shimahara} 

In Fig.\ref{tmtsfd-wave}, we can see that the pairing correlation
for $f({\bf k})=\cos(2k_x)-\cos(k_y)$ is indeed enhanced, 
while $d_{x^2-y^2}$ pairing correlation is not.   
The result provides another indication that the pairing symmetry is 
dictated by the dominant spin-fluctuation.

To summarize, the present result 
supports the spin-fluctuation mediated anisotropic pairing 
scenario in organic superconductors, by the same token as in the high $T_C$ 
cuprates discussed in \cite{KA1,KA2}.  
The enhancement of the local pair amplitude grows with $S({\bf Q})$,
while the off-diagonal long-range order ($p_{LR}$) 
stops to do so as the system approaches half-filled 
square lattice.   
It is an interesting future problem whether this 
is caused by the magnetic ordering or the metal-insulator transition, 
which is difficult to discern at present, since both of them occur 
at half-filled square lattice (at least for sufficiently large $U$).  
The possibility of the triplet pairing proposed in ref.\onlinecite{VD} also 
poses an intriguing issue.

Numerical calculations were performed at the Supercomputer Center,
Institute for Solid State Physics, University of Tokyo,
and at the Computer Center of the University of Tokyo.
K.K. acknowledges support by the Grant-in-Aid for Scientific
Research from the Ministry of Education of Japan.

%%%%%%%%%%%%%%%%%%%%  References %%%%%%%%%%%%%%%%%%%%%%%%%%%%%%%%

\end{multicols}

\begin{references}
\bibitem{incomme} Experimentally the spin density wave in (TMTSF)$_2$X is 
shown to be incommensurate in the $y$-direction (T. Takahashi {\it et al},
J. Phys. Soc. Jpn. {\bf 55}, 1364 (1986)). In the present context, this is 
related to the 
fact that the peak of $S({\bf q})$ around $(\pi/2,\pi)$ is broader in the 
$y$-direction (see the inset of Fig.\protect\ref{tmtsfd-wave}(b)), 
so that a small perturbation may change
the spin structure in that direction. Even so, 
the conclusion that the electrons in a pair is 
separated by two lattice spacings in the $x$-direction should not 
be affected.
\end{references}
\end{document}